\documentclass[11pt,showpacs,preprintnumbers,amsmath,amssymb,prd,nofootinbib,superscriptaddress]{revtex4-2}

\usepackage{dcolumn}
\usepackage{bm}
\usepackage{ifpdf}
\usepackage{hyperref}
\usepackage{xcolor,color,graphicx,graphics,physics}
\usepackage[spanish,english]{babel}
\usepackage[latin1]{inputenc}
\usepackage[OT1]{fontenc}
\usepackage{latexsym,amssymb,amsmath,amsfonts, slashed,cancel, simpler-wick}
\usepackage{makeidx}
\usepackage{epsfig,subfigure}
\usepackage{natbib}
\usepackage{epstopdf}
\usepackage{mathrsfs}
\usepackage{hyperref}
\hypersetup{colorlinks=true, linkcolor=blue, citecolor=blue, urlcolor=blue}
\usepackage{enumerate}
\usepackage{tikz}
\usepackage{feynmp-auto}  
\usepackage{tikz-feynman}
\tikzfeynmanset{compat=1.1.0}
\usepackage{fixmath}

\everymath{\displaystyle}
\usepackage{graphicx}

\usepackage[T1]{fontenc}
\usepackage{amsmath}
\usepackage{amssymb}
\usepackage{graphicx}
\usepackage{xcolor}

\newcommand{\bea}{\begin{eqnarray}}
\newcommand{\eea}{\end{eqnarray}}

\newcommand{\orcid}[1]{\href{https://orcid.org/#1}{\includegraphics[width=10pt]{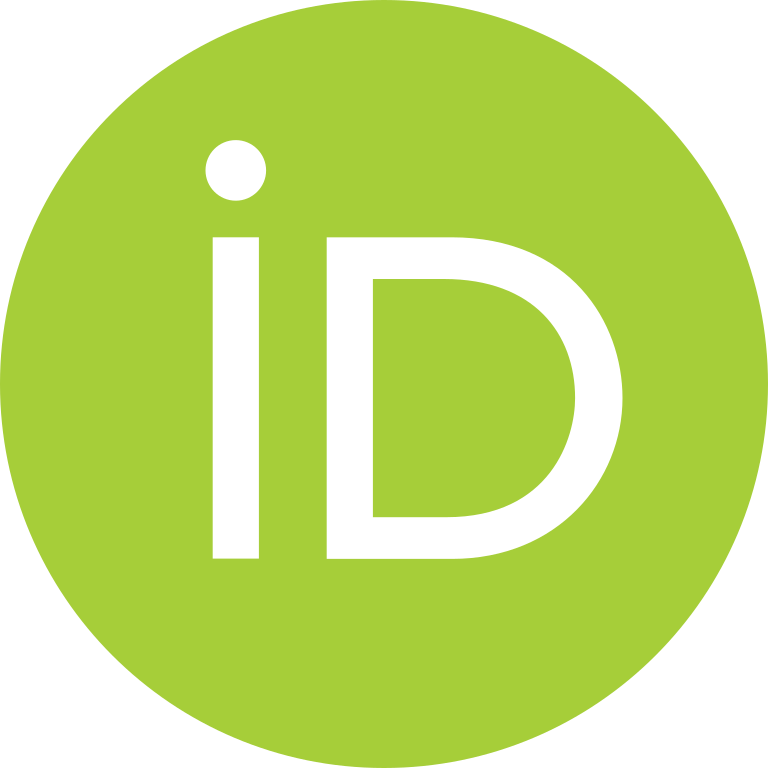}}}

\begin{document}

\title{Diffeomorphism-Like Symmetry in Gravitoelectromagnetism}

\author{L. A. S. Evangelista \orcid{0009-0002-3136-2234}}
\email{lucassouza@fisica.ufmt.br}
\affiliation{Programa de P\'{o}s-Gradua\c{c}\~{a}o em F\'{\i}sica, Instituto de F\'{\i}sica,\\ 
Universidade Federal de Mato Grosso, Cuiab\'{a}, Brasil}

\author{A. F. Santos \orcid{0000-0002-2505-5273}}
\email{alesandroferreira@fisica.ufmt.br}
\affiliation{Programa de P\'{o}s-Gradua\c{c}\~{a}o em F\'{\i}sica, Instituto de F\'{\i}sica,\\ 
Universidade Federal de Mato Grosso, Cuiab\'{a}, Brasil}

\begin{abstract}

Gravitoelectromagnetism in the Weyl formalism is investigated through an analysis of the consequences of a restricted gauge symmetry acting on the tensor field $A_{\mu\nu}$. The propagator associated with the GEM field is explicitly derived and decomposed within the Barnes--Rivers formalism, revealing contributions from the spin-2, spin-1, and scalar spin-0 sectors. By coupling the theory to conserved sources, it is shown that only the spin-2 and scalar sectors contribute to physical processes, while the spin-1 component decouples. The resulting effective propagator can then be written in a compact metric form closely resembling the graviton propagator of linearized General Relativity. The role of gauge fixing is also analyzed by considering both Lorentz-like and de Donder-type gauge conditions. It is shown that the Lorentz-like gauge is consistent with the restricted gauge symmetry of the theory and leads to gauge-independent physical amplitudes, whereas the de Donder gauge introduces residual gauge-dependent scalar contributions, signaling an incompatibility with the underlying symmetry structure. The gauge symmetry is further extended to the fermionic and electromagnetic sectors through diffeomorphism-like transformations. In both cases, conserved currents are derived and shown to coincide with the corresponding energy-momentum tensors, implying that the GEM field couples to matter in the same manner as in linearized gravity. Finally, the associated Ward identities are verified, providing a nontrivial consistency check of the gauge structure and interaction vertices of the theory.

\end{abstract}

\maketitle

\section{Introduction}

The investigation of possible connections between gravity and electromagnetism dates back to the end of the nineteenth century, when O.~Heaviside attempted to formulate Maxwell-like equations for gravity in order to describe gravitational phenomena within a framework analogous to electromagnetism \cite{heaviside1893gravitational}. Although this approach did not lead to a fully consistent theory, it established a line of research that motivated subsequent attempts to explore analogies and possible unifications between these fundamental interactions. Throughout the twentieth century, several proposals were developed with this objective, both in geometrical formulations and within the framework of quantum field theory.

One of the earliest well-structured attempts at unification was proposed within the Kaluza--Klein framework, in which gravity and electromagnetism are described in a five-dimensional space-time \cite{kaluza1921unitatsproblem, klein1926quantentheorie}. In this approach, the introduction of an extra dimension allows the electromagnetic field to emerge naturally as part of the higher-dimensional space-time metric. Although no experimental confirmation has been obtained, this formulation established an important conceptual foundation for later developments, particularly in supergravity and string theory \cite{freedman1985hidden}.

Another important approach was developed by H.~Weyl, who proposed a generalization of General Relativity (GR) based on a local gauge symmetry associated with scale transformations \cite{weyl1929gravitation}. Within this formalism, parallel transport does not preserve the length of vectors, allowing their magnitudes to vary from point to point in space-time. As a consequence, a vector field naturally emerges and can be identified with the electromagnetic potential. Although the theory was later abandoned due to physical inconsistencies, it played a fundamental role in the development of the modern concept of gauge symmetry by establishing the idea that fundamental interactions may be described in terms of local symmetries.

Within the framework of quantum field theory, the description of gravity requires the formulation of a theory in flat space-time capable of consistently describing a massless spin-2 field. In this context, the Fierz--Pauli formalism provides one of the most direct starting points \cite{fierz1939relativistic}. In this approach, gravity is described by a symmetric rank-2 tensor field, $h_{\mu\nu}$, propagating in Minkowski space-time, with dynamics governed by a quadratic Lagrangian constructed to eliminate unphysical degrees of freedom. The theory exhibits a gauge symmetry associated with linear transformations analogous to linearized diffeomorphisms, ensuring that only the physical transverse components propagate. Furthermore, the consistent coupling to external sources naturally leads to the appearance of the energy-momentum tensor as the quantity that couples universally to the spin-2 field.

However, in field-theoretic approaches formulated in flat space-time, the structure of the gauge symmetry plays a central role in determining the physical content of the theory. In particular, depending on the gauge choice, additional propagating modes beyond the spin-2 sector may emerge, raising important questions regarding their physical interpretation and consistency \cite{de2014massive}. In this context, the identification of the truly physical degrees of freedom and the elimination of unphysical contributions become essential aspects in the construction of a consistent gravitational theory.

A complementary perspective is provided by Gravitoelectromagnetism (GEM), in which gravity in the weak-field regime is described in close analogy with Maxwellian electromagnetism \cite{faqir2010lagrangian, bakopoulos2016gravitoelectromagnetism, mashhoon2003gravitoelectromagnetism, costa2008gravitoelectromagnetic,caceres2025gravitoelectromagnetismkerrschildweyldouble}. Within this framework, a formulation based on the decomposition of the Weyl tensor separates the gravitational field into gravitoelectric and gravitomagnetic components, allowing gravitational dynamics to be reinterpreted in terms of quantities analogous to those appearing in electromagnetic theory \cite{ramos2006derivation, ramos2006differential, ramos2018weyl, ramos2020abelian}. From this construction, an effective Lagrangian for a spin-2 field propagating in flat space-time can be obtained, providing a suitable framework for quantization. In this context, the gravitational interaction is described by a massless tensor field, $A_{\mu\nu}$, whose structure resembles that of gauge theories such as QED, although involving higher-rank tensor objects. Unlike the Fierz--Pauli formalism, this field is not introduced as a perturbation of the metric, but instead emerges naturally from the underlying structure of the theory.

Despite these structural advantages, several aspects of the theory remain not fully understood. In particular, the restrictive gauge condition imposed on the field $A_{\mu\nu}$ \cite{ramos2020abelian}, which is required to preserve the analogy with electrodynamics, leads to the appearance of additional propagating modes in the corresponding propagator. In the present work, the propagator associated with this field is derived, and it is shown that, although such modes arise at the intermediate level, only the spin-2 sector contributes to physical processes, such as scattering amplitudes. It is further demonstrated that an appropriate gauge-fixing procedure eliminates these extra modes, thereby restoring the physical consistency of the theory.

Finally, the gauge symmetry structure associated with the field $A_{\mu\nu}$ is analyzed. This symmetry takes the form of a diffeomorphism-like transformation and can be consistently extended to the fermionic and electromagnetic sectors, leading to effective couplings that reproduce the interaction pattern of linearized GR \cite{choi1993lowest}. It is important to emphasize, however, that the symmetry considered here does not correspond to a complete realization of space-time diffeomorphism invariance. Instead, it should be understood as a restricted effective symmetry implemented in flat space-time, whose purpose is to investigate how characteristic gravitational coupling structures, such as the emergence of the energy-momentum tensor, may arise from a controlled gauge construction. In this sense, the objective is not to reproduce GR itself, but rather to explore how a constrained symmetry structure can encode features typically associated with gravitational interactions.

This work is organized as follows. In Section \ref{SectionII}, the fundamental theoretical aspects of GEM are presented. Section \ref{SectionIII} is devoted to the analysis of the propagator structure. In Subsection \ref{SubsectionA}, the propagator is explicitly derived, while in Subsection \ref{SubsectionB} a simple physical process is analyzed in order to identify the spin components that contribute to the interaction. In Subsection \ref{SubsectionC}, a gauge-fixing Lagrangian is introduced to eliminate the additional spin modes. In Section \ref{SectionIV}, the gauge structure associated with the coupled fields is investigated, and the corresponding energy-momentum tensors for the fermionic and electromagnetic sectors are derived in Subsections \ref{SubsectionA2} and \ref{SubsectionB2}, respectively. Finally, the conclusions are presented in Section \ref{SectionV}.

\section{Theoretical concepts of GEM}\label{SectionII}

This section is devoted to the discussion of the fundamental aspects of Gravitoelectromagnetism, a theoretical framework that seeks to describe gravitational phenomena through a mathematical analogy with electromagnetism. In general, three main approaches are employed in the formulation of GEM: (i) the analogy between GR in the weak-field limit and Maxwell equations \cite{mashhoon2003gravitoelectromagnetism}, known as linearized GEM; (ii) the decomposition of the Weyl curvature tensor into its gravitoelectric and gravitomagnetic components \cite{faqir2010lagrangian}, commonly referred to as Weyl GEM; and (iii) an analysis based on the tidal tensor formalism \cite{costa2008gravitoelectromagnetic}.

In the present work, the second approach is adopted, since it admits a well-established Lagrangian formulation that is particularly suitable for field quantization \cite{alesandrogravitacional, faqir2010lagrangian}. Within this formalism, the Weyl curvature tensor is defined as
\begin{eqnarray}
	C_{\alpha\sigma\mu\nu} &=& R_{\alpha\sigma\mu\nu} - \frac{1}{2} \left( R_{\nu\alpha} g_{\mu\sigma} + R_{\mu\sigma} g_{\nu\alpha} - R_{\nu\sigma} g_{\mu\alpha} - R_{\mu\alpha} g_{\nu\sigma} \right)+ \frac{1}{6} R \left( g_{\nu\alpha} g_{\mu\sigma} - g_{\nu\sigma} g_{\mu\alpha} \right),
\end{eqnarray}
where $R_{\alpha\sigma\mu\nu}$ is the Riemann tensor, $R_{\nu\alpha}$ denotes the Ricci tensor, $R$ is the Ricci scalar, and $g_{\mu\alpha}$ is the metric tensor. The Weyl tensor can be decomposed into its gravitoelectric and gravitomagnetic components, given by
\begin{align}
	{E}_{ij} &= -C_{0i0j}, \\
	{B}_{ij} &= \frac{1}{2} \epsilon_{ikl} C^{kl}_{\phantom{kl}0j}, 
\end{align}
with $i,j,k,\ldots = 1,2,3$. It is important to emphasize that these fields are more generally defined as $E_{\mu\nu}=u^\kappa u^\lambda C_{\mu\kappa\nu\lambda}$ and $B_{\mu\nu}=\frac{1}{2} u^\kappa u^\lambda \epsilon_{\alpha\rho\kappa\mu} C^{\alpha\rho}{}_{\nu\lambda}$, respectively, where $u^\lambda$ denotes the four-velocity of the observer \cite{Campbell, Ber}. In a local rest frame associated with this four-velocity, only the spatial components of these tensors are nonvanishing. Furthermore, since both tensors are trace-free, they together account for the ten independent components of the Weyl tensor. This decomposition provides a natural framework to analyze gravitational tidal effects and spacetime curvature in close analogy with the electric and magnetic fields in electromagnetism.

The GEM, or Maxwell-like, equations in flat spacetime in the presence of sources \cite{ramos2006derivation} are given by
\begin{align}
	\partial^{i} {E}^{ij} &= 4\pi G \rho^j, \\
	\partial^{i} {B}^{ij} &= 0, \\
	\epsilon^{\langle ikl \rangle} \partial^k {B}^{lj} - \frac{1}{c} \frac{\partial {E}^{ij}}{\partial t} &= \frac{4\pi G}{c} J^{ij}, \\
	\epsilon^{\langle ikl \rangle} \partial^k {E}^{lj} + \frac{1}{c} \frac{\partial {B}^{ij}}{\partial t} &= 0,
\end{align}
where $\rho^j$ denotes the mass density vector, $J^{ij}$ is the mass current density tensor, $c$ is the speed of light, and $G$ is the gravitational constant. Within the Weyl GEM formalism, these fields encode the dynamical degrees of freedom of the gravitational interaction, in close analogy with the electric and magnetic fields in electromagnetism.

The fields $E^{ij}$ and $B^{ij}$ are symmetric tensors and can be expressed in terms of potentials as
\begin{align}
	E &= - \mathrm{grad}\,\varphi - \frac{1}{c}\frac{\partial \tilde{A}}{\partial t},\\
	B &= \mathrm{curl}\,\tilde{A},
\end{align}
where $\varphi$ denotes the scalar potential, analogous to the electromagnetic scalar potential, and $\tilde{A}$ represents a tensor potential with components $A^{ij}$. These quantities generalize the electromagnetic potentials by carrying an additional tensorial index structure. The operators $\mathrm{grad}$ and $\mathrm{curl}$ are defined as generalizations of the standard differential operators, constructed through Clebsch-Gordan techniques in order to consistently act on higher-rank tensor fields \cite{ramos2006differential}.

With these definitions, one can introduce the GEM field strength tensor,
\begin{eqnarray}
	F^{\mu\nu\alpha} = \partial^\mu A^{\nu\alpha} - \partial^\nu A^{\mu\alpha}, \label{tensorfield}
\end{eqnarray}
and its dual tensor,
\begin{eqnarray}
	G^{\mu\nu\alpha} = \frac{1}{2}\epsilon^{\mu\nu\gamma\sigma} \eta^{\alpha\xi} F_{\gamma\sigma\xi}.
\end{eqnarray}
The GEM field equations can then be written in covariant form as
\begin{eqnarray}
	\partial_{\mu} F^{\mu\nu\alpha} &=& -\frac{4\pi G}{c}\mathcal{J}^{\nu\alpha}, \label{tensorF}\\
	\partial_{\mu} G^{\mu\nu\alpha} &=& 0, \label{tensorG}
\end{eqnarray}
where $\mathcal{J}^{\nu\alpha}$ is a second-rank tensor constructed from the mass density vector $\rho^j$ and the mass current tensor $J^{ij}$. From these equations, the GEM Lagrangian density is defined as
\begin{eqnarray}
	\mathcal{L}_{\text{GEM}} = -\frac{1}{16\pi} F_{\mu\nu\alpha} F^{\mu\nu\alpha} + G \mathcal{J}^{\nu\alpha} A_{\nu\alpha}.
\end{eqnarray}

A comparison is now established between this approach, in which the tensor potential $A_{\mu\nu}$ is treated as the fundamental field, and the linearized Einstein equations formulated in terms of the metric perturbation $h_{\mu\nu}$. In contrast to the latter formulation, which is based on a perturbative expansion of the metric around flat space-time, the present framework employs $A_{\mu\nu}$ as the primary dynamical variable, thereby providing an alternative description of the gravitational interaction.

Combining Eq.~(\ref{tensorfield}) with Eq.~(\ref{tensorF}), one obtains the GEM field equation in the presence of sources,
\begin{eqnarray}
	\Box A^{\nu\alpha} - \partial^\nu(\partial_\mu A^{\mu\alpha}) = 4\pi G \mathcal{J}^{\nu\alpha}.
\end{eqnarray}
Due to the gauge invariance of the GEM potential, analogous to the electromagnetic case, one can impose a Lorentz-type gauge condition $\partial_\mu A^{\mu\alpha} = 0$, which reduces the field equation to
\begin{eqnarray}
	\Box A^{\nu\alpha} = 4\pi G \mathcal{J}^{\nu\alpha}.
\end{eqnarray}
This expression closely resembles the linearized Einstein field equations in terms of the trace-reversed metric perturbation,
\begin{eqnarray}
	\Box \bar{h}_{\mu\nu} = 16\pi G T_{\mu\nu},
\end{eqnarray}
where $\bar{h}_{\mu\nu} = h_{\mu\nu} - \frac{1}{2}\eta_{\mu\nu} h$ \cite{misner1973k}. The structural similarity between these equations highlights the correspondence between GEM and GR. In particular, the tensor $A^{\nu\alpha}$ in the Weyl GEM formalism encodes the essential dynamical content of the theory, playing a role analogous to that of the metric perturbation in the linearized regime.

An important distinction, however, is that the tensor potential $A_{\mu\nu}$ arises naturally within the Weyl GEM construction, rather than being introduced as a perturbation of the metric. This feature provides a conceptually distinct framework, in which the gravitational interaction is described in terms of tensorial gauge fields. Furthermore, the existence of a well-defined Lagrangian formulation makes this approach especially suitable for quantization, allowing for a consistent interpretation in terms of a spin-2 field.

For a more detailed discussion of the differences between GEM and GR, see Refs.~\cite{bakopoulos2016gravitoelectromagnetism, mashhoon2003gravitoelectromagnetism, alesandrogravitacional}. In the next section, the formalism developed here is employed to analyze the propagator associated with $A^{\mu\nu}$. In particular, the manner in which the gauge symmetry affects its structure is investigated, and the propagator is expressed systematically and transparently through the use of transverse and longitudinal projectors.

\section{Propagator Structure of the $A_{\mu\nu}$ Field}\label{SectionIII}

This section is devoted to the derivation of the propagator within the Weyl GEM formalism. It is shown that, in order to preserve a structure analogous to electromagnetism, the adopted gauge symmetry leads to a propagator containing additional spin modes. Nevertheless, it is demonstrated that, in physical processes, only the spin-2 sector contributes, while the extra modes do not correspond to physical propagating degrees of freedom.

\subsection{Derivation of the Propagator}\label{SubsectionA}
The free GEM Lagrangian is given by
\begin{align}
	\mathcal{L}_{\text{G}}
	=
	-\frac{1}{16\pi}
	F_{\mu\nu\alpha} F^{\mu\nu\alpha}.
	\label{GEMField}
\end{align}
The GEM field transforms under a gauge symmetry according to
\begin{equation}
	A_{\mu\nu}
	\longrightarrow
	A_{\mu\nu}
	+
	\partial_\mu \theta_\nu.
\end{equation}
In general, the symmetry of the tensor field would require a symmetrized transformation. Instead of introducing this symmetrization explicitly, the following condition is imposed:
\begin{align}
	\partial_\mu \theta_\nu = \partial_\nu \theta_\mu,
\end{align}
which implies that $\theta_\nu$ can be written as a pure gradient,
\begin{equation}
	\theta_\nu = \partial_\nu \lambda,
\end{equation}
where $\lambda$ is a scalar function. This condition ensures the symmetry of the tensor field while restricting the number of independent gauge degrees of freedom. It also establishes a formal analogy with electrodynamics, in which the gauge symmetry is controlled by a scalar parameter. As a consequence, the resulting theory does not correspond to a fully general spin-2 gauge theory, but rather to a tensorial theory with an effectively induced scalar gauge symmetry. The gauge parameter effectively selects a longitudinal sector of the symmetry. This restriction plays a central role in ensuring the consistency of the model and distinguishes it from full diffeomorphism invariance.

One may ask why not consider the usual gauge transformation of linearized GR, which leads to diffeomorphism invariance. In other words, why not use
\begin{align}
	A_{\mu\nu}\to A_{\mu\nu}+\partial_\mu\xi_\nu+\partial_\nu\xi_\mu, \label{hgauge}
\end{align}
where $\xi_\mu$ represents a vector gauge parameter. If this transformation is applied, the resulting Lagrangian is not invariant under it, indicating that the Weyl GEM formulation is not diffeomorphism invariant. To restore such invariance, the GEM Lagrangian would need to be reformulated, effectively moving toward the GR regime and requiring a reconstruction of the entire Weyl GEM structure, which would deviate from the original objective of constructing a formalism analogous to electromagnetism.

It is important to emphasize that GEM is a weak-field theory, operating in a quasi-static regime. Therefore, its domain of validity is limited. Even though it does not constitute a fundamental theory of gravity, the Weyl GEM formalism provides an effective and useful framework within its range of applicability.

Considering only the contraction of the field strength tensor in Eq.~\eqref{GEMField}, we obtain
\begin{align}
	F_{\mu\nu\alpha}F^{\mu\nu\alpha}
	&=
	2\left(
	\partial_\mu A_{\nu\alpha} \partial^\mu A^{\nu\alpha}
	-\partial_\mu A_{\nu\alpha} \partial^\nu A^{\mu\alpha}
	\right).
\end{align}
Substituting this result into the free Lagrangian, we find
\begin{align}
	\mathcal{L}_G
	=
	\frac{1}{8\pi}
	A_{\mu\nu}
	\left(
	\eta^{\mu\rho}\Box - \partial^\mu \partial^\rho
	\right)
	A_{\rho}{}^{\nu},
\end{align}
which leads to the corresponding field equation
\begin{align}
	\Box A^{\rho \sigma} - \partial^\rho \partial_\mu A^{\mu\sigma} = 0.
\end{align}
In momentum space, with the replacement $\partial_\mu \to i k_\mu$, the GEM Lagrangian takes the form
\begin{align}
	\mathcal{L}_{G}
	=
	\frac{1}{8\pi}
	A^{\mu\nu}
	O_{\mu\nu,\rho\sigma}(k)
	A^{\rho\sigma},
\end{align}
where $O_{\mu\nu,\rho\sigma}(k)$ is the kinetic operator, given by
\begin{align}
	O_{\mu\nu,\rho\sigma}(k)
	=
	k^2 \eta_{\mu\rho}\eta_{\nu\sigma}
	-
	k_\mu k_\rho \eta_{\nu\sigma}.\label{bilinearl}
\end{align}
Although the propagator could be obtained directly from this operator, such an approach would obscure its underlying spin structure. In order to make this structure explicit, transverse and longitudinal projectors are introduced,
\begin{align}
	\theta_{\mu\nu}
	=
	\eta_{\mu\nu}
	-
	\frac{k_\mu k_\nu}{k^2},
	\qquad
	\omega_{\mu\nu}
	=
	\frac{k_\mu k_\nu}{k^2},\label{transpropertie}
\end{align}
which satisfy the completeness relation
\begin{align}
	\theta_{\mu\nu} + \omega_{\mu\nu} = \eta_{\mu\nu}.
	\label{spinpropertie}
\end{align}
In terms of these projectors, the kinetic operator can be written as
\begin{align}
	O_{\mu\nu,\rho \sigma}
	=
	k^2 \theta_{\mu\rho}\eta_{\nu\sigma}.
	\label{kineticaloperator}
\end{align}
This operator is not symmetric under the exchange of indices. However, since the GEM field itself is symmetric, only the symmetric part of the operator contributes to the quadratic action. Recalling that a symmetric operator can be constructed as
\begin{align}
	O_{\mu\nu,\rho\sigma}^{(\text{sym})}
	=
	\frac{1}{2}
	\left(
	O_{\mu\nu,\rho\sigma}
	+
	O_{\nu\mu,\rho\sigma}
	\right),
\end{align}
the fully symmetrized operator in both index pairs is obtained,
\begin{align}
	O_{\mu\nu,\rho\sigma}^{(\text{full})}
	=
	\frac{k^2}{4}
	\left(
	\theta_{\mu\rho}\eta_{\nu\sigma}
	+
	\theta_{\nu\rho}\eta_{\mu\sigma}
	+
	\theta_{\mu\sigma}\eta_{\nu\rho}
	+
	\theta_{\nu\sigma}\eta_{\mu\rho}
	\right).
\end{align}
Using the relation \eqref{spinpropertie}, this expression can be rewritten as
\begin{align}
	O_{\mu\nu,\rho\sigma}^{(\text{full})}
	=
	\frac{k^2}{4}
	\Big(
	\theta_{\mu\rho}\theta_{\nu\sigma}
	+
	\theta_{\mu\rho}\omega_{\nu\sigma}
	+
	\theta_{\nu\rho}\theta_{\mu\sigma}
	+
	\theta_{\nu\rho}\omega_{\mu\sigma}
	+
	\theta_{\mu\sigma}\theta_{\nu\rho}
	+
	\theta_{\mu\sigma}\omega_{\nu\rho}
	+
	\theta_{\nu\sigma}\theta_{\mu\rho}
	+
	\theta_{\nu\sigma}\omega_{\mu\rho}
	\Big).
\end{align}
It is important to emphasize that the symmetrized operator preserves the original structure of the GEM Lagrangian. In other words, the symmetrization procedure does not alter the dynamical content of the theory, but merely ensures consistency with the symmetry properties of the tensor field.

In order to analyze the spin content propagated by the field, the Barnes--Rivers decomposition is employed, allowing the kinetic operator associated with a symmetric rank-2 tensor to be expressed in terms of spin projection operators \cite{nunes1993extending, accioly2002propagator}. These projectors are given by
\begin{align}
	P^{(2)}_{\mu\nu,\rho\sigma}&=\frac{1}{2}\left(\theta_{\mu\rho}\theta_{\nu\sigma}+\theta_{\mu\sigma}\theta_{\nu\rho}\right)-\frac{1}{3}\theta_{\mu\nu}\theta_{\rho\sigma},\\
	P^{(1)}_{\mu\nu,\rho\sigma}&=\frac{1}{2}\left(\theta_{\mu\rho}\omega_{\nu\sigma}+\theta_{\mu\sigma}\omega_{\nu\rho}+\theta_{\nu\rho}\omega_{\mu\sigma}+\theta_{\nu\sigma}\omega_{\mu\rho}\right),\\
	P^{(0s)}_{\mu\nu,\rho\sigma}&=\frac{1}{3}\theta_{\mu\nu}\theta_{\rho\sigma},\\
	P^{(0l)}_{\mu\nu,\rho\sigma}&=\omega_{\mu\nu}\omega_{\rho\sigma},\\
	P^{(0m)}_{\mu\nu,\rho\sigma}&=\theta_{\mu\nu}\omega_{\rho\sigma}+\omega_{\mu\nu}\theta_{\rho\sigma},
\end{align}
where the superscripts $(0,1,2)$ denote the corresponding spin sectors, while $s$, $l$, and $m$ refer to the scalar, longitudinal scalar, and mixed components, respectively. These operators satisfy the completeness relation
\begin{align}
	\left[P^{(2)}+P^{(1)}+P^{(0s)}+P^{(0l)}\right]_{\mu\nu,\rho\sigma}
	=\frac{1}{2}\left(\eta_{\mu\rho}\eta_{\nu\sigma}+\eta_{\mu\sigma}\eta_{\nu\rho}\right)
	\equiv I_{\mu\nu,\rho\sigma},
\end{align}
where $I_{\mu\nu,\rho\sigma}$ is the identity operator in the space of symmetric rank-2 tensors.

Using these projectors, and identifying the contribution of each spin sector, the kinetic operator can be written as
\begin{align}
	O_{\mu\nu,\rho\sigma}^{(full)}
	= k^2\left(P^{(2)}_{\mu\nu,\rho\sigma}
	+ P^{(0s)}_{\mu\nu,\rho\sigma}
	+ \frac{1}{2} P^{(1)}_{\mu\nu,\rho\sigma}\right),\label{originalkinetial}
\end{align}
where the longitudinal and mixed scalar sectors do not contribute explicitly to the physical dynamics.

By definition, the propagator is required to satisfy the relation
\begin{align}
	O_{\mu\nu,\alpha\beta}^{(full)} D^{\alpha\beta}{}_{\rho\sigma}
	= I_{\mu\nu,\rho\sigma}
	= \left[P^{(2)}+P^{(1)}+P^{(0s)}+P^{(0l)}\right]_{\mu\nu,\rho\sigma}.
	\label{ODrelation}
\end{align}
The following Ansatz is then considered
\begin{align}
	D_{\mu\nu,\rho\sigma}
	= a\, P^{(2)}_{\mu\nu,\rho\sigma}
	+ b\, P^{(1)}_{\mu\nu,\rho\sigma}
	+ c\, P^{(0s)}_{\mu\nu,\rho\sigma}
	+ d\, P^{(0l)}_{\mu\nu,\rho\sigma}.
\end{align}
Substituting into Eq.~\eqref{ODrelation}, we obtain
\begin{align}
	a=\frac{1}{k^2}, \qquad
	b=\frac{2}{k^2}, \qquad
	c=\frac{1}{k^2},
\end{align}
while the longitudinal coefficient remains undetermined, reflecting its pure gauge character. Consequently, one is free to choose $d=0$. With this choice, the propagator for the GEM theory is given by
\begin{align}
	D_{\mu\nu,\rho\sigma}
	= \frac{1}{k^2}\left(P^{(2)} + 2 P^{(1)} + P^{(0s)}\right).
	\label{propagator}
\end{align}
This result shows that the GEM field propagates spin-2, spin-1, and spin-0 modes. Consequently, the theory does not describe a purely spin-2 field, as additional lower-spin contributions arise from the gauge structure of the model. However, when the propagator is coupled to physical sources, one can explicitly verify which sectors effectively contribute to observable processes, typically isolating the physically relevant degrees of freedom.

\subsection{Spin Content in Physical Processes}\label{SubsectionB}

The previous results can now be employed to analyze which spin sectors effectively contribute to the GEM field in the presence of sources. This analysis can be carried out through the construction of the generating functional associated with the transition amplitude \cite{peskin2018introduction}. Accordingly, one writes
\begin{align}
	Z[\mathcal{J}] = \int \mathcal{D}A \;\exp\left[i \int d^4x \, \mathcal{L} \right],
	\label{functional}
\end{align}
where $\mathcal{D}A$ denotes the functional integration over all field configurations.

The GEM Lagrangian in the presence of an external source is given by
\begin{align}
	\mathcal{L}_{GEM} = -\frac{1}{16\pi} F_{\mu\nu\alpha} F^{\mu\nu\alpha} + \mathcal{J}_{\mu\nu} A^{\mu\nu}.
\end{align}
In momentum space, and in terms of the kinetic operator, it can be written as
\begin{align}
	\mathcal{L}_{GEM} = \frac{1}{2} A^{\mu\nu} O_{\mu\nu,\rho\sigma} A^{\rho\sigma}+ \mathcal{J}_{\mu\nu} A^{\mu\nu}.
\end{align}
Substituting into Eq.~\eqref{functional}, we obtain
\begin{align}
	Z[\mathcal{J}] = \int \mathcal{D}A \;\exp\left[i \int d^4x \left( \frac{1}{2} A O A + \mathcal{J} A \right) \right],
\end{align}
where tensor indices have been suppressed for notational simplicity.

Completing the square by defining the shifted field
\begin{align}
	A' = A + O^{-1} \mathcal{J},
\end{align}
we obtain
\begin{align}
	Z[\mathcal{J}]
	&= \int \mathcal{D}A \;\exp\left\{ i \int d^4x \left[ \frac{1}{2} \left(A' - O^{-1}\mathcal{J}\right) O \left(A' - O^{-1}\mathcal{J}\right) + \mathcal{J}\left(A' - O^{-1}\mathcal{J}\right) \right] \right\} \nonumber \\
	&= \int \mathcal{D}A \;\exp\left[ \frac{i}{2} \int d^4x \left( A' O A' - \mathcal{J} O^{-1} \mathcal{J} \right) \right].
\end{align}
The functional integral over $\mathcal{D}A$ contributes only as an overall normalization constant. Using $O^{-1} = D$, where $D_{\mu\nu,\rho\sigma}$ is the propagator, the generating functional reduces to
\begin{align}
	Z[\mathcal{J}] = Z[0] \exp\left( -\frac{i}{2} \mathcal{J}^{\mu\nu} D_{\mu\nu,\rho\sigma} \mathcal{J}^{\rho\sigma} \right),
\end{align}
where $Z[0]$ is the generating functional in the absence of sources. It is important to note that only the interaction contribution is being considered here. Furthermore, the space-time integration is implicitly incorporated into the functional contraction, since the analysis is performed in momentum space.

The tree-level amplitude associated with the exchange of a single quantum is therefore given by
\begin{align}
	\mathcal{M} \sim \mathcal{J}^{\mu\nu} D_{\mu\nu,\rho\sigma} \mathcal{J}^{\rho\sigma}.
\end{align}

It has been shown that the GEM propagator contains contributions from the spin-2, spin-1, and spin-0 sectors. Upon coupling it to the previous expression, one obtains
\begin{align}
	\mathcal{M}=\frac{1}{k^2}\mathcal{J}^{\mu\nu}\left(P^{(2)}_{\mu\nu,\rho\sigma}+2P^{(1)}_{\mu\nu,\rho\sigma}+P^{(0s)}_{\mu\nu,\rho\sigma}\right)\mathcal{J}^{\rho\sigma}.
\end{align}
Since $k_\mu \mathcal{J}^{\mu\nu}=0$, it follows that $\omega_{\mu\nu}\mathcal{J}^{\mu\nu}=0$ and $\theta_{\mu\nu}\mathcal{J}^{\mu\nu}=\mathcal{J}_{\mu}{}^\mu$. By analyzing each projector separately, the spin-2 contribution is first obtained as
\begin{align}
	\mathcal{J}^{\mu\nu}P^{(2)}\mathcal{J}^{\rho\sigma}&=\mathcal{J}^{\mu\nu}\left[\frac 12\left(\theta_{\mu\rho} \theta_{\nu\sigma}+\theta_{\mu\sigma}\theta_{\nu\rho}\right)-\frac 13 \theta_{\mu\nu} \theta_{\rho\sigma}\right]\mathcal{J}^{\rho\sigma}\nonumber\\
	&=\mathcal{J}^{\mu\nu}\mathcal{J}_{\mu\nu}-\frac 13\mathcal{J}_{\nu}{}^\nu \mathcal{J}_{\sigma}{}^{\sigma}.\label{spin2part}
\end{align}
For spin-1, we have
\begin{align}
	\mathcal{J}^{\mu\nu}P^{(1)}\mathcal{J}^{\rho\sigma}&=\mathcal{J}^{\mu\nu}\left[\frac 12\left(\theta_{\mu\rho} \omega_{\nu\sigma}+\theta_{\mu\sigma}\omega_{\nu\rho}+\theta_{\nu\rho}\omega_{\mu\sigma}+\theta_{\nu\sigma}\omega_{\mu\rho}\right)\right]\mathcal{J}^{\rho\sigma}\nonumber\\
	&=0.
\end{align}
Finally, for spin-0,
\begin{align}
	\mathcal{J}^{\mu\nu}P^{(0s)}\mathcal{J}^{\rho\sigma}&=\mathcal{J}^{\mu\nu}\left(\frac 13 \theta_{\mu\nu}\theta_{\rho\sigma}\right)\mathcal{J}^{\rho\sigma}\nonumber\\
	&=\frac 13 \mathcal{J}_{\nu}{}^\nu \mathcal{J}_{\sigma}{}^\sigma.
\end{align}
By summing all contributions, the following expression is obtained
\begin{align}
	\mathcal{M}=\mathcal{J}^{\mu\nu}D_{\mu\nu,\rho\sigma}\mathcal{J}^{\rho\sigma}=\mathcal{J}^{\mu\nu}\mathcal{J}_{\mu\nu}.
\end{align}
It should be noted that, for the effective interaction with conserved sources, no contribution arises from the spin-1 sector. Furthermore, although a spin-0 contribution is present, it is exactly canceled by part of the spin-2 sector. Consequently, the GEM theory does not correspond kinematically to a purely spin-2 theory, although its interaction with conserved sources occurs effectively as if it did. Therefore, when coupled to physical conserved sources, the theory reproduces an effective spin-2 dynamics and can be consistently employed in the computation of physical processes within this regime.

However, it may be questioned whether these additional contributions correspond merely to gauge artifacts, which would disappear after the inclusion of an appropriate gauge-fixing Lagrangian in the total action. In the next subsection, a gauge-fixing Lagrangian is introduced, and the extent to which these additional modes contribute to the final propagator is investigated.

\subsection{Gauge Fixing and Consistency}\label{SubsectionC}

This section is devoted to the introduction of a gauge-fixing Lagrangian into the total action. The objective is to eliminate the additional propagating modes. As discussed in the previous subsection, the spin-0 contribution is canceled by part of the spin-2 sector when the theory is coupled to conserved sources, as shown in Eq.~\eqref{spin2part}, leading to an effectively spin-2 dynamics in physical processes. The purpose here is therefore to remove the extra propagating modes while preserving the complete spin-2 structure of the theory.

Since a Lorentz-like gauge condition, $\partial_\mu A^{\mu\nu}=0$, is to be imposed, a gauge-fixing Lagrangian is introduced in the form
\begin{align}
	\mathcal{L}_{gf}=-\frac{1}{2\xi}\left(\partial_\mu A^{\mu\nu}\right)\left(\partial^\rho A_{\rho\nu}\right).
\end{align}
Following a structure analogous to that of Eq.~\eqref{bilinearl}, the previous expression can be rewritten as
\begin{align}
	\mathcal{L}_{gf}=\frac{1}{2}A^{\mu\nu}\mathcal{O}^{(gf)}_{\mu\nu,\rho\sigma}A^{\rho\sigma},
\end{align} 
where the kinetic operator associated with the gauge-fixing term is
\begin{align}
	\mathcal{O}^{(gf)}_{\mu\nu,\rho\sigma}=\frac 1\xi\partial_\mu \partial_\rho \eta_{\nu\sigma}.
\end{align}
It should be noted that this operator is not symmetric and therefore must be symmetrized in order to remain consistent with the symmetric structure of the GEM field. The fully symmetrized kinetic operator in momentum space can then be written as
\begin{align}
	\mathcal{O}^{(gf,full)}_{\mu\nu,\rho\sigma}=\frac{1}{4\xi}\left(k_\mu k_\rho\eta_{\nu\sigma}+k_\nu k_\rho \eta_{\mu\sigma}+k_\mu k_\sigma \eta_{\nu\rho}+k_\nu k_\sigma \eta_{\mu\rho}\right).
\end{align}
By introducing the projectors defined in Eq.~\eqref{transpropertie} and employing the properties given in Eq.~\eqref{spinpropertie}, the corresponding spin projectors in the Barnes--Rivers basis can be identified, leading to the following kinetic operator:
\begin{align}
	\mathcal{O}^{(gf,full)}_{\mu\nu,\rho\sigma}=\frac{k^2}{\xi}\left(\frac 12 P^{(1)}_{\mu\nu,\rho\sigma}+P^{(0l)}_{\mu\nu,\rho\sigma}\right).
\end{align}
By combining this operator with the original one given in Eq.~\eqref{originalkinetial}, the total propagator associated with the gravitoelectromagnetic field can be written as
\begin{align} 
	D^{(total)}_{\mu\nu,\rho\sigma}(k)=\frac{1}{k^2}\left[P^{(2)}_{\mu\nu,\rho\sigma}+P^{(0s)}_{\mu\nu,\rho\sigma}+\frac{2\xi}{(1+\xi)}P^{(1)}_{\mu\nu,\rho\sigma}+\xi P^{(0l)}_{\mu\nu,\rho\sigma}\right].\label{propgaugee}
\end{align}
The propagator exhibits a single massless pole at $k^2=0$. Although multiple spin components appear in the residue, only the transverse components contribute when coupled to conserved currents. Note that the gauge-fixing structure gives rise to a longitudinal scalar contribution, as well as the already present spin-1 term. Although the scalar projector $P^{(0s)}$ remains in the propagator, it does not correspond to an independent physical contribution. Indeed, when contracted with a conserved current, the combination $P^{(2)} + P^{(0s)}$ reduces to the symmetric tensor structure
\begin{align}
	P^{(2)}_{\mu\nu,\rho\sigma} + P^{(0s)}_{\mu\nu,\rho\sigma} = \frac{1}{2}(\theta_{\mu\rho}\theta_{\nu\sigma}+\theta_{\mu\sigma}\theta_{\nu\rho}),
\end{align}
so that the physical amplitudes depend only on the contraction $\mathcal{J}^{\mu\nu}\mathcal{J}_{\mu\nu}$, which, as previously discussed, reproduces an effective spin-2 structure. Therefore, the scalar sector does not propagate as an independent degree of freedom in physical processes. Furthermore, the additional longitudinal scalar contribution does not contribute to physical amplitudes, since it involves longitudinal projectors and thus vanishes upon contraction with conserved currents satisfying $k_\mu \mathcal{J}^{\mu\nu}=0$. In addition, the inclusion of the gauge-fixing Lagrangian shifts the spin-1 contribution into a gauge-dependent sector, which vanishes in the limit $\xi=0$. Hence, it is suppressed by an appropriate gauge choice in physical observables. Consequently, the GEM theory, together with the Lorentz-like gauge-fixing contribution, yields a propagator that effectively contributes only through the spin-2 sector, independently of the gauge parameter, as expected.

In the calculation of physical processes, Eq.~\eqref{propgaugee}, for $\xi=0$, can be written in metric form as
\begin{align}
	D^{(total)}_{\mu\nu,\rho\sigma}(k)=\frac{1}{2k^2}\left(\eta_{\mu\rho}\eta_{\nu\sigma}+ \eta_{\mu\sigma}\eta_{\nu\rho}\right),\label{finalpropagator}
\end{align}
where the following relations have been used \cite{accioly2002propagator}
\begin{align}
	P^{(2)}_{\mu\nu,\rho\sigma}&=\frac{1}{2}\left(\eta_{\mu\rho}\eta_{\nu\sigma}+ \eta_{\mu\sigma}\eta_{\nu\rho}\right)-\frac 13 \eta_{\mu\nu}\eta_{\rho\sigma}-\left(P^{(1)}_{\mu\nu,\rho\sigma}+\frac 23 P^{(0l)}_{\mu\nu,\rho\sigma}-\frac 13 P^{(0m)}_{\mu\nu,\rho\sigma}\right),\\
	P^{(0s)}_{\mu\nu,\rho\sigma}&=\frac 13 \eta_{\mu\nu}\eta_{\rho\sigma}-\frac 13 \left(P^{(0l)}_{\mu\nu,\rho\sigma}+P^{(0m)}_{\mu\nu,\rho\sigma}\right).
\end{align}

It is important to note that the final representation of the propagator relevant for physical processes, obtained in Eq.~\eqref{finalpropagator}, differs from the structure commonly employed in Weyl GEM scattering applications \cite{faqir2010lagrangian, alesandrogravitacional, evangelista2026gravitational}. In those works, the graviton propagator associated with the linearized GR regime is adopted as an effective description under the assumption of full diffeomorphism invariance. It should be emphasized that such applications may be understood as resulting from different assumptions concerning the underlying gauge structure. Nevertheless, although a different propagator is employed in those analyses, the corresponding physical predictions remain unchanged when the propagator derived in the present work is used instead. Therefore, the expression obtained here should be regarded as the natural propagator associated with the present theory.

For completeness, a de Donder-like gauge may also be considered, corresponding to the standard gauge-fixing condition employed in GR. In this case, the following condition is imposed:
\begin{align}
	\partial_\mu A^{\mu\nu}-\frac 12 \partial^\nu A=0,
\end{align}
where $A=A^\mu {}_\mu$. Accordingly, the gauge-fixing Lagrangian is defined as
\begin{align}
	\mathcal{L}_{gf}=-\frac{1}{2\xi}\left(\partial^\mu A_{\mu\nu} -\frac 12 \partial_\nu A\right)\left(\partial^\rho A_\rho{}^\nu -\frac 12 \partial^\nu A\right).
\end{align}
This leads to the following kinetic operator for the gauge-fixing term,
\begin{align}
	\mathcal{O}_{\mu\nu,\rho\sigma}^{(gf,full)}=\frac {1}{2\xi} \left[\frac 12\left(\partial_\mu \partial_\rho \eta_{\nu\sigma}+\partial_\nu \partial_\rho \eta_{\mu\sigma}+\partial_\mu \partial_\sigma \eta_{\nu\rho}+\partial_\nu\partial_\sigma \eta_{\mu\rho}-\eta_{\mu\nu} \eta_{\rho\sigma}\Box\right)+ \partial_\mu\partial_\nu \eta_{\rho\sigma}+\partial_\rho \partial_\sigma \eta_{\mu\nu}\right].
\end{align}
After transforming to momentum space and identifying the corresponding spin projectors, the following expression is obtained
\begin{align}
	\mathcal{O}_{\mu\nu,\rho\sigma}^{(gf,full)}=\frac{k^2}{\xi}\left(\frac{1}{2}P^{(1)}_{\mu\nu,\rho\sigma}-\frac{3}{4}P^{(0s)}_{\mu\nu,\rho\sigma}+\frac{11}{4}P^{(0l)}_{\mu\nu,\rho\sigma}+\frac 14 P^{(0m)}_{\mu\nu,\rho\sigma}\right).\label{Operator69}
\end{align}
This operator is expressed in the Barnes--Rivers basis $\left\{P^{(2)},P^{(1)},P^{(0s)},P^{(0l)},P^{(0m)}\right\}$. Since this structure contains mixing projectors involving scalar contributions, including both transverse and longitudinal components, the inversion of the kinetic operator is no longer diagonal, in contrast to the previous case. It is therefore assumed that the operator can be written as
\begin{align}
	O=x_2 P^{(2)}+x_1 P^{(1)}+x_0^{(s)} P^{(0s)}+ x_0^{(l)}P^{(0l)}+ x_0^{(m)}P^{(0m)},
\end{align}
where each coefficient $x_i$ multiplies the corresponding projector, the inverse operator takes the form \cite{nunes1993extending}
\begin{align}
	D=\frac{1}{x_2}P^{(2)}+\frac{1}{x_1}P^{(1)}+ \frac{1}{x_0^{(s)} x_0^{(l)}-3(x_0^{(m)})^2}\left(x_0^{(l)} P^{(0s)}+x_0^{(s)}P^{(0l)}-x_0^{(m)} P^{(0m)}\right).
\end{align}
By combining Eq.~\eqref{Operator69} with the original kinetic operator in the absence of gauge fixing, the total propagator is obtained as
\begin{align}
	D_{\mu\nu,\rho\sigma}^{(total)}(k)=\frac{1}{k^2}\left[P^{(2)}+\frac{2\xi}{\xi+1}P^{(1)}+\frac{\xi}{11\xi-9}\left(11 P^{(0s)}+(4\xi-3)P^{(0l)}-P^{(0m)}\right)\right].
\end{align}

An interesting feature arises when a de Donder-like gauge condition is imposed within the present framework. In this case, the resulting propagator exhibits a nontrivial dependence on the gauge parameter in the scalar sector, as can be seen from the coefficient multiplying the $P^{(0s)}$ projector.

At first sight, this behavior may suggest the presence of a gauge-dependent physical contribution, which would indicate a possible inconsistency of the theory. However, this feature should instead be understood as a consequence of the fact that the underlying gauge symmetry of the model does not correspond to a complete realization of diffeomorphism invariance. As a result, gauge conditions motivated by linearized GR are not necessarily compatible with the intrinsic structure of the present framework.

This indicates that the de Donder gauge is not well suited to the present theory, since it mixes spin sectors in a manner that does not properly reflect the physical degrees of freedom of the model. In contrast, the Lorentz-like gauge condition adopted in this work leads to a consistent propagator structure, in which gauge-dependent contributions do not affect physical observables.

In the next section, the determination of the conserved current arising from gauge symmetry is discussed for both the Dirac and electromagnetic fields. Once the fermionic current is obtained from gauge invariance, the corresponding Ward identities are analyzed. These identities provide a nontrivial consistency check of the theory, ensuring that the interaction vertex remains compatible with the underlying gauge symmetry at the level of physical amplitudes.

\section{Gauge Symmetry and the Structure of Interaction Currents}\label{SectionIV}

In this section, the role of gauge symmetry in determining the interaction structure of the theory is investigated. By analyzing the transformation properties of the GEM and matter fields, the form of the conserved current associated with the symmetry is derived. It is then shown that this current satisfies the corresponding Ward identities, thereby ensuring the consistency of the interaction vertex and the decoupling of unphysical degrees of freedom.

\subsection{Fermionic Sector}\label{SubsectionA2}

Let us begin by recalling the full GEM Lagrangian, given by
\begin{align}
	\mathcal{L}_{\text{GEM}} = -\frac{1}{16\pi} F_{\mu\nu\alpha} F^{\mu\nu\alpha} + \mathcal{J}^{\mu\nu} A_{\mu\nu}.
\end{align}
The GEM potential tensor transforms under the gauge symmetry according to
\begin{align}
	\delta A_{\mu\nu} = \kappa\partial_\mu \theta_\nu,
\end{align}
where $\kappa$ denotes the GEM coupling constant.

Although the gauge symmetry considered here is restricted to the longitudinal sector, it is sufficient to ensure the consistency of the theory at the level of physical observables. This feature is nontrivial, since the underlying theory propagates multiple spin sectors, as previously discussed. Nevertheless, only the spin-2 contribution survives in physical processes, leading to an effective coupling structure analogous to that of linearized gravity.

The discussion of the gauge structure is now extended to include the Dirac field. The full Lagrangian is given by
\begin{align}
	\mathcal{L}_{tot} = -\frac{1}{16\pi} F_{\mu\nu\alpha} F^{\mu\nu\alpha}
	+ \mathcal{J}^{\mu\nu} A_{\mu\nu}
	+ \overline{\psi}\left(i \gamma^\mu \overleftrightarrow{\partial_\mu} - m \right)\psi,
	\label{fulllagrangian}
\end{align}
where the Dirac sector is written in its symmetrized form.

The Dirac field is assumed to transform, to first order in the gauge parameter, according to an analysis analogous to that employed in GR in the weak-field limit \cite{weyl1929gravitation}. Explicitly, one has
\begin{align}
	\delta \psi &= - \kappa \theta^\rho \partial_\rho \psi, \nonumber\\
	\delta \overline{\psi} &= - \kappa \theta^\rho \partial_\rho \overline{\psi}
	\label{psitransformation}
\end{align}
This transformation should be interpreted as an effective realization of a diffeomorphism-like symmetry, restricted to a specific sector of field configurations. It does not correspond to a genuine general coordinate transformation, but instead reproduces, in a controlled setting, its algebraic action on the fields.

The manner in which each field transforms under the gauge symmetry is now analyzed. The discussion begins with the GEM sector. Under the gauge transformation, the variation of the Lagrangian is given by
\begin{align}
	\delta \mathcal{L}_{\text{GEM}}
	&= -\frac{1}{16\pi} (\delta F_{\mu\nu\alpha}) F^{\mu\nu\alpha}
	- \frac{1}{16\pi} F_{\mu\nu\alpha} (\delta F^{\mu\nu\alpha})
	+ \mathcal{J}^{\mu\nu} \delta A_{\mu\nu}.
\end{align}
By employing the definition of the field strength tensor, the following expression is obtained
\begin{align}
	\delta F_{\mu\nu\alpha}
	= \partial_\mu (\delta A_{\nu\alpha}) - \partial_\nu (\delta A_{\mu\alpha}).
\end{align}
After substituting the gauge transformation $\delta A_{\mu\nu} = \partial_\mu \theta_\nu$, it is found
\begin{align}
	\delta F_{\mu\nu\alpha}
	= \kappa\partial_\mu \partial_\nu \theta_\alpha - \kappa\partial_\nu \partial_\mu \theta_\alpha.
\end{align}
By using the commutativity of partial derivatives, $\partial_\mu \partial_\nu = \partial_\nu \partial_\mu$, one obtains
\begin{align}
	\delta F_{\mu\nu\alpha} = 0.
\end{align}
Therefore, the variation of the kinetic term vanishes identically, leaving
\begin{align}
	\delta \mathcal{L}_{\text{GEM}} = \kappa\partial_\mu \theta_\nu\mathcal{J}^{\mu\nu} .
	\label{gemgauge}
\end{align}
It is important to note that, although the full GEM Lagrangian is not invariant under the gauge transformation due to the presence of the source term, the free Lagrangian remains invariant.

The analysis is now extended to the Dirac sector. By considering only this contribution in Eq.~\eqref{fulllagrangian}, one obtains
\begin{align}
	\delta\mathcal{L}_D
	&= (\delta\overline{\psi})(i\gamma^\mu \overleftrightarrow{\partial_\mu}-m)(\delta\psi) \nonumber \\
	&= \frac{i}{2} \left[(\delta\overline{\psi})\gamma^\mu \partial_\mu (\delta\psi) - \partial_\mu(\delta\overline{\psi})\gamma^\mu(\delta\psi)\right] - m(\delta\overline{\psi})(\delta\psi).
\end{align}
After substituting the gauge transformations for $\psi$ and $\overline{\psi}$, the following expression is obtained
\begin{align}
	\delta\mathcal{L}_D
	&=\frac{i\kappa}{2}\big[-\theta^\rho(\partial_\rho\overline{\psi})\gamma^\mu (\partial_\mu \psi)-\overline{\psi}\gamma^\mu(\partial_\mu\theta^\rho)(\partial_\rho\psi)-\overline{\psi}\gamma^\mu \theta^\rho (\partial_\mu \partial_\rho \psi) \nonumber\\
	&\quad+(\partial_\mu \theta^\rho) (\partial_\rho \overline{\psi})\gamma^\mu \psi
	+\theta^\rho(\partial_\mu \partial_\rho \overline{\psi})\gamma^\mu \psi+(\partial_\mu \overline{\psi})\gamma^\mu \theta^\rho(\partial_\rho \psi)\big] \nonumber\\
	&\quad -m\kappa\left[-\theta^\rho \partial_\rho \overline{\psi}\psi-\overline{\psi}\theta^\rho\partial_\rho\psi\right].
\end{align}
The first term of the mass contribution can be written as
\begin{align}
	\theta^\rho \partial_\rho \overline{\psi}\psi 
	= \theta^\rho \partial_\rho (\overline{\psi}\psi)- \overline{\psi}\theta^\rho \partial_\rho \psi.
\end{align}
By separating the terms proportional to $\theta^\rho$ and $\partial_\mu\theta^{\rho}$, the following result is derived
\begin{align}
	\delta \mathcal{L}_D&=\frac{i\kappa}{2}\big\{\theta^\rho\big[-(\partial_\rho\overline{\psi})\gamma^\mu (\partial_\mu \psi)-\overline{\psi}\gamma^\mu(\partial_\mu\partial_\rho \psi)+(\partial_\mu \partial_\rho \overline{\psi})\gamma^\mu \psi+(\partial_\mu \overline{\psi})\gamma^\mu(\partial_\rho\psi)\nonumber\\
	&\quad +m\partial_\rho(\overline{\psi}\psi)\big]+\partial_\mu \theta^\rho\left[-\overline{\psi}\gamma^\mu(\partial_\rho\psi)+(\partial_\rho\overline{\psi})\gamma^\mu \psi\right]\big\}.
\end{align}
The $\theta^\rho$ contribution can be written as $\theta^\rho \partial_\rho \mathcal{C}$, where $\mathcal{C}$ represents the expression inside the brackets without the derivative acting on it. Consequently,
\begin{align}
	\delta \mathcal{L}_D&=\frac{i\kappa}{2}\left\{\theta^\rho \partial_\rho\left[-\overline{\psi}\gamma^\mu(\partial_\mu\psi)+(\partial_\mu\overline{\psi})\gamma^\mu \psi+m\overline{\psi}\psi\right]+\partial_\mu\theta^\rho\left[-\overline{\psi}\gamma^\mu(\partial_\rho\psi)+(\partial_\rho\overline{\psi})\gamma^\mu\psi\right]\right\}.
\end{align}
Using the identity $\theta^\rho\partial_\rho\mathcal{C}=\partial_\rho(\theta^\rho \mathcal{C})-\partial_\rho\theta^\rho \mathcal{C}$ and neglecting the total derivative contribution, the variation reduces to
\begin{align}
	\delta \mathcal{L}_D&=\frac{i\kappa}{2}\left\{-\partial_\rho \theta^\rho \left[-\overline{\psi}\gamma^\mu(\partial_\mu\psi)+(\partial_\mu\overline{\psi})\gamma^\mu \psi+m\overline{\psi}\psi\right]+\partial_\mu\theta^\rho\left[-\overline{\psi}\gamma^\mu(\partial_\rho\psi)+(\partial_\rho\overline{\psi})\gamma^\mu\psi\right]\right\}.
\end{align}
The first term is recognized as the Dirac Lagrangian. After reorganizing the indices, the expression can be written as
\begin{align}
	\delta\mathcal{L}_D=\kappa\partial_\mu\theta_\nu\left[-\frac{i}{2}\left(\overline{\psi}\gamma^\mu\partial^\nu\psi-\partial^\nu\overline{\psi}\gamma^\mu\psi\right)+\eta^{\mu\nu}\mathcal{L}_D\right].\label{eq87}
\end{align}
Furthermore, it can be symmetrized as
\begin{align}
	-\frac{i\kappa}{2}\partial_\mu\theta_\nu\left(\overline{\psi}\gamma^\mu\partial^\nu\psi-\partial^\nu\overline{\psi}\gamma^\mu\psi\right)
	&=-\frac{i\kappa}{4}\partial_\mu\theta_\nu\left(\overline{\psi}\gamma^\mu\partial^\nu\psi-\partial^\nu\overline{\psi}\gamma^\mu\psi\right)\nonumber\\
	&\quad-\frac{i\kappa}{4}\partial_\nu\theta_\mu\left(\overline{\psi}\gamma^\nu\partial^\mu\psi-\partial^\mu\overline{\psi}\gamma^\nu\psi\right).
\end{align}
Using the condition $\partial_\mu\theta_\nu=\partial_\nu\theta_\mu$, we find
\begin{align}
	-\frac{i\kappa}{2}\partial_\mu\theta_\nu\left(\overline{\psi}\gamma^\mu\partial^\nu\psi-\partial^\nu\overline{\psi}\gamma^\mu\psi\right)
	=-\frac{i\kappa}{4}\partial_\mu\theta_\nu\big(\overline{\psi}\gamma^\mu\partial^\nu\psi-\partial^\nu\overline{\psi}\gamma^\mu\psi+\overline{\psi}\gamma^\nu\partial^\mu\psi-\partial^\mu\overline{\psi}\gamma^\nu\psi\big).
\end{align}
Using the bidirectional derivative notation, we obtain
\begin{align}
	-\frac{i\kappa}{2}\partial_\mu\theta_\nu\left(\overline{\psi}\gamma^\mu\partial^\nu\psi-\partial^\nu\overline{\psi}\gamma^\mu\psi\right)
	=-\frac{i\kappa}{4}\partial_\mu\theta_\nu\big(\overline{\psi}\gamma^\mu\overleftrightarrow{\partial^\nu}\psi+\overline{\psi}\gamma^\nu\overleftrightarrow{\partial^\mu}\psi\big).\label{symmetrizedequation}
\end{align}
Substituting this result into Eq.~\eqref{eq87}, the fully transformed Dirac Lagrangian is written as
\begin{align}
	\delta\mathcal{L}_D=-\frac{i\kappa}{4}\partial_\mu\theta_\nu\big(\overline{\psi}\gamma^\mu\overleftrightarrow{\partial^\nu}\psi+\overline{\psi}\gamma^\nu\overleftrightarrow{\partial^\mu}\psi\big)+\kappa\partial_\mu\theta_\nu\eta^{\mu\nu}\mathcal{L}_D.\label{eq91}
\end{align}

By collecting the transformed expressions given in Eq.~\eqref{gemgauge} and Eq.~\eqref{eq91}, and imposing the condition $\delta \mathcal{L}_{tot}=0$, the following expression is obtained
\begin{align}
	-\frac{i\kappa}{4}\partial_\mu\theta_\nu\big(\overline{\psi}\gamma^\mu\overleftrightarrow{\partial^\nu}\psi+\overline{\psi}\gamma^\nu\overleftrightarrow{\partial^\mu}\psi\big)+\kappa\partial_\mu\theta_\nu\eta^{\mu\nu}\mathcal{L}_D+\kappa\partial_\mu\theta_\nu\mathcal{J}^{\mu\nu}=0.
\end{align}

From this structure, the conserved current tensor can be identified as
\begin{align}
	\mathcal{J}^{\mu\nu}_D=\frac{i}{4}\big(\overline{\psi}\gamma^\mu\overleftrightarrow{\partial^\nu}\psi+\overline{\psi}\gamma^\nu\overleftrightarrow{\partial^\mu}\psi\big)-\eta^{\mu\nu}\mathcal{L}_D,
\end{align}
where the subscript $D$ indicates that this corresponds to the Dirac fermionic current.
This expression reproduces the structure of the symmetrized Belinfante-Rosenfeld (BR) energy-momentum tensor \cite{acevedo2023teoremas},
\begin{align}
	T^{\mu\nu}_{BR}=\frac{i}{4}\overline{\psi}\left(\gamma^\mu \overleftrightarrow{\partial^\nu} +\gamma^\nu \overleftrightarrow{\partial^\mu} \right)\psi -\eta^{\mu\nu}\mathcal{L}_D.\label{BRTensor}
\end{align}
Therefore, one can consistently identify
\begin{align}
	\mathcal{J}^{\mu\nu}_D=T^{\mu\nu}_{BR}\label{relation JT}.
\end{align}

The gauge transformation employed in \eqref{psitransformation} captures only the transport part of a diffeomorphism-like transformation. As a consequence, the variation of the Lagrangian density determines the conserved current in the form
\begin{equation}
	\mathcal{L}' = \mathcal{L} + (\partial_\mu \theta_\nu)\,\mathcal{J}^{\mu\nu}.
\end{equation}

However, this transformation does not fully implement the behavior of the Lagrangian as a scalar density under spacetime transformations and, consequently, no spin connection contribution is generated. A complete diffeomorphism-like transformation could in principle be introduced within the formalism of quantum fields in curved spacetime. Nevertheless, since such an extension lies beyond the scope of the present work, it will not be pursued further here.

The energy-momentum tensor obtained in \eqref{relation JT} is the same object that appears in the construction of Feynman vertices in linearized GR \cite{faqir2010lagrangian}. This structural equivalence indicates that the GEM field couples to matter in the same way as in GR \cite{hollik2014quantum}, namely
\begin{align}
	\mathcal{L}_{\text{int}}=\kappa A_{\mu\nu}T^{\mu\nu}.
\end{align}
Therefore, the interaction Lagrangian between the GEM field and fermions is given by
\begin{align}
	\mathcal{L}_{A\psi\psi}=\frac{i\kappa}{4}A_{\mu\nu}\overline{\psi}\left(\gamma^\mu \overleftrightarrow{\partial^\nu} +\gamma^\nu \overleftrightarrow{\partial^\mu} \right)\psi.
\end{align}
As a consequence, the interaction vertex takes the form
\begin{align}
	V^{\mu\nu}_{A\psi\psi}=\frac{i\kappa}{4}\left[\gamma^\mu\left(p_1+p_2\right)^\nu+\gamma^\nu\left(p_1+p_2\right)^\mu\right],
\end{align}
where $p_1$ and $p_2$ denote the incoming and outgoing fermion momenta, respectively.

For fermionic fields, the contribution proportional to $\eta^{\mu\nu}\mathcal{L}_D$ is proportional to the equations of motion and therefore vanishes in on-shell matrix elements. As a result, the interaction vertex is entirely determined by the kinetic part of the energy-momentum tensor. Consequently, although the trace term won't be explicitly generated in the construction, the resulting current will remains conserved and satisfies the corresponding Ward identities at the level of physical amplitudes as we will see further in the work.

It is important to note that the vertex obtained here differs from the structure presented in \cite{faqir2010lagrangian, choi1993lowest}. However, in those references, the vertex does not satisfy the Ward identity. Therefore, in order to construct a current that fulfills this condition, it is necessary to adopt a symmetrized form, as performed in the present analysis, which leads to a consistent interaction vertex. Without the symmetrization procedure introduced in Eq.~\eqref{symmetrizedequation}, one would recover the same vertex structure reported in those references.

Instead of verifying the Ward identity directly at the level of the vertex function, which may require cumbersome manipulations in momentum space, the conservation of the corresponding current may equivalently be analyzed in coordinate space.

In the present construction, the current obtained from gauge invariance is identified with the energy-momentum tensor, i.e.,
\begin{align}
	\mathcal{J}^{\mu\nu} = T^{\mu\nu}.
\end{align}
Therefore, the Ward identity in momentum space,
\begin{align}
	q_\mu \mathcal{J}^{\mu\nu} = 0,
\end{align}
is directly related to the local conservation law
\begin{align}
	\partial_\mu T^{\mu\nu} = 0.
\end{align}

To make this connection explicit, consider the matrix element
\begin{align}
	\mathcal{J}^{\mu\nu}(q) = \int d^4x \, e^{iq\cdot x} \langle f | T^{\mu\nu}(x) | i \rangle.
\end{align}
Acting with $q_\mu$ yields
\begin{align}
	q_\mu \mathcal{J}^{\mu\nu}(q)
	=
	i \int d^4x \, e^{iq\cdot x} \langle f | \partial_\mu T^{\mu\nu}(x) | i \rangle,
\end{align}
where integration by parts has been used, assuming that boundary terms vanish.

Thus, the conservation of the energy-momentum tensor in coordinate space,
\begin{align}
	\partial_\mu T^{\mu\nu} = 0,
\end{align}
implies the Ward identity in momentum space,
\begin{align}
	q_\mu \mathcal{J}^{\mu\nu} = 0,
\end{align}
for physical matrix elements.

In this way, the verification of the Ward identity can be consistently carried out at the level of local conservation laws, avoiding the need for explicit evaluation of vertex contractions in momentum space. This procedure is completely analogous to the standard derivation of Ward identities in gauge theories, where current conservation in coordinate space ensures the corresponding constraints in momentum space.

With these considerations established, the quantity $\partial_\mu \mathcal{J}^{\mu\nu}$ is now computed for the fermionic sector. The energy-momentum tensor may be written in the expanded form
\begin{align}
	\mathcal{J}^{\mu\nu}_D=\frac {i\kappa}{4}\left(\overline{\psi}\gamma^\mu \partial^\nu \psi-\partial^\nu \overline{\psi} \gamma^\mu \psi+ \overline{\psi}\gamma^\nu \partial^\mu \psi-\partial^\mu \overline{\psi}\gamma^\nu\psi\right).
\end{align}
Using the field equations,
\begin{align}
	\gamma^\mu \partial_\mu\psi=-i m\psi,\quad\quad\partial_\mu \overline{\psi}\gamma^\mu=i m \overline{\psi},\quad\quad\Box \psi=-m^2\psi,\quad\quad\Box\overline{\psi}=-m^2\overline{\psi},
\end{align}
and substituting them into the expression above, one finds that the terms cancel pairwise, leading to
\begin{align}
	\partial_\mu \mathcal{J}^{\mu\nu}_D=0.
\end{align}
Hence, it has been shown that, even under a restricted gauge transformation for the GEM field, the transformation of the fermionic sector gives rise to a transport current analogous to that obtained in linearized GR. As a consequence, the GEM field couples to matter in a manner analogous to linearized GR, while simultaneously satisfying a conserved current structure, which in turn guarantees the validity of the corresponding Ward identity for the interaction vertex.

\subsection{Electromagnetic Sector}\label{SubsectionB2}

The gauge symmetry analysis is now extended to the electromagnetic field. The electromagnetic Lagrangian is given by
\begin{align}
	\mathcal{L}_{EM}=-\frac 14 \mathcal{F}_{\mu\nu}\mathcal{F}^{\mu\nu},
\end{align}
where $\mathcal{F}_{\mu\nu}=\partial_\mu \mathcal{A}_\nu-\partial_\nu \mathcal{A}_\mu$. For the vector field, a diffeomorphism-like transformation is adopted \cite{wald2010general}, namely,
\begin{align}
	\delta \mathcal{A}_\mu=\kappa\left(\theta^\rho\partial_\rho \mathcal{A}_\mu+\partial_\mu \theta^\rho \mathcal{A}_\rho\right).
\end{align}

The total Lagrangian is given by
\begin{align}
	\mathcal{L}_{tot} = -\frac{1}{16\pi} F_{\mu\nu\alpha} F^{\mu\nu\alpha}
	+ \mathcal{J}^{\mu\nu} A_{\mu\nu}-\frac 14 \mathcal{F}_{\mu\nu} \mathcal{F}^{\mu\nu}.
\end{align}

Since the gauge transformation of the GEM field has already been analyzed, only the transformation of the electromagnetic sector remains to be considered. Thus,
\begin{align}
	\delta \mathcal{L}_{EM}
	&=-\frac 14 (\delta \mathcal{F}_{\mu\nu})\mathcal{F}^{\mu\nu}
	-\frac 14 \mathcal{F}_{\mu\nu}(\delta \mathcal{F}^{\mu\nu}).
\end{align}
Using the definition of the field strength tensor, we have
\begin{align}
	\delta \mathcal{F}_{\mu\nu}
	=\partial_\mu(\delta \mathcal{A}_\nu)-\partial_\nu(\delta \mathcal{A}_\mu).
\end{align}
Substituting the transformation, we obtain
\begin{align}
	\delta \mathcal{F}_{\mu\nu}
	&=\kappa\Big[\partial_\mu\left(\theta^\rho\partial_\rho \mathcal{A}_\nu+\partial_\nu \theta^\rho \mathcal{A}_\rho\right)
	-\partial_\nu\left(\theta^\rho\partial_\rho \mathcal{A}_\mu+\partial_\mu \theta^\rho \mathcal{A}_\rho\right)\Big].
\end{align}
Expanding the derivatives, one finds
\begin{align}
	\delta \mathcal{F}_{\mu\nu}
	&=\kappa\Big[(\partial_\mu\theta^\rho)\partial_\rho \mathcal{A}_\nu+\theta^\rho \partial_\mu\partial_\rho \mathcal{A}_\nu
	+(\partial_\mu\partial_\nu\theta^\rho)\mathcal{A}_\rho+(\partial_\nu\theta^\rho)\partial_\mu \mathcal{A}_\rho \nonumber\\
	&\quad-(\partial_\nu\theta^\rho)\partial_\rho \mathcal{A}_\mu-\theta^\rho \partial_\nu\partial_\rho \mathcal{A}_\mu
	-(\partial_\nu\partial_\mu\theta^\rho)\mathcal{A}_\rho-(\partial_\mu\theta^\rho)\partial_\nu \mathcal{A}_\rho\Big].
\end{align}
Using the commutativity of partial derivatives and rearranging terms, this expression reduces to
\begin{align}
	\delta \mathcal{F}_{\mu\nu}
	=\kappa\left[\theta^\rho \partial_\rho \mathcal{F}_{\mu\nu}+(\partial_\mu\theta^\rho)\mathcal{F}_{\rho\nu}+(\partial_\nu\theta^\rho)\mathcal{F}_{\mu\rho}\right].
\end{align}
Substituting into the variation of the Lagrangian and keeping terms up to first order, we obtain
\begin{align}
	\delta \mathcal{L}_{EM}
	=(\partial_\mu \theta_\nu)\left[\mathcal{F}^{\mu\lambda}\mathcal{F}^\nu{}_\lambda-\frac 14 \eta^{\mu\nu}\mathcal{F}_{\alpha\beta}\mathcal{F}^{\alpha\beta}\right].
\end{align}
This leads to the current
\begin{align}
	\mathcal{J}^{\mu\nu}_{EM}=-\mathcal{F}^{\mu\lambda}\mathcal{F}^\nu{}_\lambda+\frac 14 \eta^{\mu\nu}\mathcal{F}_{\alpha\beta}\mathcal{F}^{\alpha\beta}.\label{currenphoton}
\end{align}

This expression coincides with the energy-momentum tensor of the electromagnetic field. Consequently, the current couples in the same manner as the electromagnetic energy-momentum tensor. Then
\begin{align}
	\mathcal{J}^{\mu\nu}_{EM}=T^{\mu\nu}_{EM}.
\end{align}

Unlike in the fermionic case, the trace contribution in the electromagnetic energy-momentum tensor does not vanish on-shell and therefore cannot be neglected. Its presence is essential to guarantee the correct symmetry, conservation, and gauge-invariant structure of the theory. In particular, this term plays a crucial role in reproducing the standard GEM-photon-photon vertex found in linearized GR.

Therefore, the emergence of the full energy-momentum tensor in the electromagnetic sector is not accidental, but rather a direct consequence of the transformation properties assigned to $\mathcal{A}_\mu$, which are sufficiently rich to encode the complete coupling structure required for consistency with gravitational interactions.

With these results, the interaction Lagrangian between the $A_{\mu\nu}$ field and photons can be written as
\begin{align}
	\mathcal{L}_{A\gamma\gamma}=\kappa A_{\mu\nu}\left(-\mathcal{F}^{\mu\lambda}\mathcal{F}^\nu{}_\lambda+\frac 14\eta^{\mu\nu}\mathcal{F}_{\alpha\beta}\mathcal{F}^{\alpha\beta}\right).
\end{align}
Such a Lagrangian leads to the following vertex:
\begin{align}
	V^{\mu\nu,\alpha\beta}_{A\gamma\gamma}(k_1,k_2) = -i\kappa \Big[ \eta^{\alpha\beta} k_1^\mu k_2^\nu -\eta^{\nu\beta} k_1^\mu k_2^\alpha -\eta^{\mu\alpha}k_1^\beta k_2^\nu+\eta^{\mu\alpha}\eta^{\nu\beta}(k_1\cdot k_2)-\frac 12 \eta^{\mu\nu}\big(\eta^{\alpha\beta}(k_1\cdot k_2)-k_1^\beta k_2^\alpha\big) \Big],
\end{align}
where $k_1$ and $k_2$ are the photon momenta. The last two expressions correspond to the Lagrangian interaction and the vertex of the linearized GR with $A_{\mu\nu}$ instead of $h_{\mu\nu}$ \cite{faqir2010lagrangian,choi1993lowest}.

With the relevant quantities properly defined, the conservation of the current in Eq.~\eqref{currenphoton}, namely $\partial_\mu\mathcal{J}^{\mu\nu}_{EM}$, can now be analyzed. To this end, the following quantity must be computed
\begin{align}
	\partial_\mu \mathcal{J}^{\mu\nu}_{EM}=\partial_\mu\left(-\mathcal{F}^{\mu\lambda}\mathcal{F}^\nu{}_\lambda+\frac 14 \eta^{\mu\nu}\mathcal{F}_{\alpha\beta}\mathcal{F}^{\alpha\beta}\right).
\end{align}

Starting from the trace contribution, we obtain
\begin{align}
	\partial_\mu\left(\frac 14 \eta^{\mu\nu}\mathcal{F}_{\alpha\beta}\mathcal{F}^{\alpha\beta}\right)&=
	\frac 12 \mathcal{F}^{\alpha\beta}\partial^\nu \mathcal{F}_{\alpha\beta}.\label{104}
\end{align}

The kinetic term, on the other hand, can be written as
\begin{align}
	-\partial_\mu \left(\mathcal{F}^{\mu\lambda} \mathcal{F}^\nu{}_\lambda\right)
	=
	-\big(\partial_\mu \mathcal{F}^{\mu\lambda}\big)\mathcal{F}^\nu{}_\lambda
	-\mathcal{F}^{\mu\lambda}\big(\partial_\mu \mathcal{F}^\nu {}_\lambda\big).
\end{align}
Using the field equation $\partial_\mu \mathcal{F}^{\mu\lambda}=0$, we obtain
\begin{align}
	-\partial_\mu \left(\mathcal{F}^{\mu\lambda} \mathcal{F}^\nu{}_\lambda\right)
	=
	-\mathcal{F}_{\gamma\sigma}\partial^\gamma \mathcal{F}^{\nu\sigma}.
\end{align}
This term can then be substituted into the conservation expression, and the indices may be rearranged in order to maintain a coherent notation, leading to the following expression
\begin{align}
		\partial_\mu \mathcal{J}^{\mu\nu}_{EM}=-\mathcal{F}_{\mu\lambda}\partial^\mu \mathcal{F}^{\nu\lambda}+\frac 12 \mathcal{F}_{\lambda\mu}\partial^\nu \mathcal{F}^{\lambda\mu}.
\end{align}
By performing suitable manipulations, the equation assumes the form
\begin{align}
	\partial_\mu \mathcal{J}^{\mu\nu}_{EM}	&=- \frac 12 \mathcal{F}_{\mu\lambda}\left(\partial^\mu \mathcal{F}^{\nu\lambda}+\partial^\lambda \mathcal{F}^{\mu\nu}+\partial^\nu \mathcal{F}^{\lambda\mu}\right).
\end{align}
It should be noted that the terms inside the brackets correspond precisely to the Bianchi identities, i.e.,
\begin{align}
	\partial^\mu \mathcal{F}^{\nu\lambda}+\partial^\lambda \mathcal{F}^{\mu\nu}+\partial^\nu\mathcal{F}^{\lambda\mu}=0.
\end{align}
Therefore, we get that
\begin{align}
	\partial_\mu \mathcal{J}^{\mu\nu}_{EM}=0.
\end{align}

Hence, the diffeomorphism-like transformation gives rise to the complete electromagnetic energy-momentum tensor, and the corresponding current is found to satisfy the Ward identity in coordinate space.

\section{Conclusions}\label{SectionV}

In this work, the Weyl formalism of Gravitoelectromagnetism was adopted in order to analyze the theoretical implications of the gauge symmetry imposed on the $A_{\mu\nu}$ field and its associated consequences. As a result, the propagator of the GEM field was derived explicitly, and it was found that, due to the restricted gauge symmetry, additional spin modes are propagated, namely spin-2, spin-1, and scalar spin-0 components. This analysis was carried out through the Barnes--Rivers decomposition, in which symmetric rank-2 fields are expressed as combinations of transverse and longitudinal projectors, thereby allowing a clear identification of the spin content associated with the kinetic operator.

After the propagator had been obtained, a physical process was analyzed in order to determine which spin components correspond to physical propagation. It was found that only the spin-2 and spin-0 sectors contribute. Since the spin-2 projector can be decomposed into transverse and scalar components, the resulting effective physical contribution leads to a propagator of the form given in Eq.~\eqref{finalpropagator}, which closely resembles the graviton propagator in linearized GR.

In addition, a gauge-fixing analysis was performed by considering both the Lorentz-like and de Donder gauges. For the Lorentz-like gauge, it was found that the spin-1 sector, together with an additional longitudinal spin-0 contribution, is shifted into the gauge-dependent sector and vanishes in the Landau gauge limit, $\xi=0$, so that physical processes depend effectively only on the spin-2 and spin-0 contributions. In contrast, for the de Donder gauge, it was found that the spin-1, scalar spin-0, and additional longitudinal and mixed spin-0 contributions are likewise shifted into the gauge-dependent sector. However, since the scalar spin-0 contribution does not disappear from physical processes, the resulting propagator retains a dependence on the gauge parameter $\xi$, thereby signaling an unphysical behavior associated with this gauge choice.

The gauge symmetry associated with the interaction currents in both the fermionic and electromagnetic sectors was also investigated. Since the $A_{\mu\nu}$ field exhibits a symmetry analogous to diffeomorphism invariance, diffeomorphism-like transformations were adopted for both the fermionic and electromagnetic fields. In both cases, the resulting conserved currents were found to possess the same structure as the corresponding energy-momentum tensors, implying that the $A_{\mu\nu}$ field couples to matter in a manner analogous to linearized GR. Consequently, the interaction Lagrangians and the corresponding vertices coincide with those obtained within the framework of linearized GR.

Finally, it was verified that these currents satisfy the corresponding Ward identities, thereby providing a nontrivial consistency check of the gauge structure of the theory. In this way, the results obtained in the present work offer a new perspective on the gauge structure of GEM theory, clarifying how effective spin-2 dynamics and conserved interaction currents naturally emerge within the Weyl formalism.

\section*{Acknowledgments}

This work by A. F. S. is partially supported by National Council for Scientific and Technological
Development - CNPq project No. 312406/2023-1. L. A. S. E. thank CAPES for financial support.

\section*{Data Availability Statement}

No Data associated in the manuscript.

\section*{Conflicts of Interest}

No conflict of interests in this paper.


\global\long\def\link#1#2{\href{http://eudml.org/#1}{#2}}
 \global\long\def\doi#1#2{\href{http://dx.doi.org/#1}{#2}}
 \global\long\def\arXiv#1#2{\href{http://arxiv.org/abs/#1}{arXiv:#1 [#2]}}
 \global\long\def\arXivOld#1{\href{http://arxiv.org/abs/#1}{arXiv:#1}}


\end{document}